\documentclass[10pt,conference]{IEEEtran}
\IEEEoverridecommandlockouts

\usepackage{subcaption}
\usepackage{textcomp}
\usepackage{cite} 
\usepackage{stfloats}
\usepackage[switch]{lineno}
\usepackage{amsmath,amssymb,amsfonts}
\usepackage{algorithmic}
\usepackage{graphicx}
\usepackage{textcomp}
\usepackage[dvipsnames,svgnames]{xcolor}
\usepackage{booktabs}
\usepackage{longtable}
\usepackage{multirow}
\usepackage{array}
\usepackage{enumitem}
\usepackage{colortbl}
\usepackage{verbatim}
\usepackage{url}
\usepackage{tcolorbox}
\usepackage{tabularx}
\usepackage{fontawesome5}
\usepackage[linesnumbered,lined,boxed,commentsnumbered]{algorithm2e}
\usepackage{booktabs}
\usepackage{multirow}
\usepackage{listings}
 \usepackage{fancyhdr}
 \usepackage{verbatim}
 \usepackage{flushend}
 \usepackage{threeparttable}
\usepackage[switch]{lineno}
\usepackage[marginal]{footmisc}

\usepackage{pifont}
\newcommand{\datasetO}{\textsc{ObfusEval}}

\newcommand{\Space}[1]{}

\newcommand{\checked}[1]{#1}
\newcommand{\ignore}[1]{}

\newcommand{\sinc}[1]{{\color{green}{\bf $\uparrow$#1}}}
\newcommand{\sdec}[1]{{\color{red}{\bf $\boldsymbol{\downarrow}$#1}}}
\newcommand{\inc}[1]{{\color{DarkGreen}{\bf $\boldsymbol{\Uparrow}$#1}}}
\newcommand{\dec}[1]{{\color{DarkRed}{\bf $\boldsymbol{\Downarrow}$#1}}}
\newcommand{\equa}{{\bf =}}

\makeatletter
\newcommand{\linebreakand}{%
  \end{@IEEEauthorhalign}
  \hfill\mbox{}\par
  \mbox{}\hfill\begin{@IEEEauthorhalign}
}
\makeatother

\begin{document}

\title{Unseen Horizons: Unveiling the Real Capability of LLM Code Generation Beyond the Familiar}

\author{
    \IEEEauthorblockN{
        Yuanliang Zhang\IEEEauthorrefmark{1}, 
        Yifan Xie\IEEEauthorrefmark{1}, 
        Shanshan Li\IEEEauthorrefmark{2}, 
        Ke Liu, 
        Chong Wang, 
        Zhouyang Jia, 
        Xiangbing Huang, \\
        Jie Song,
        Chaopeng Luo, 
        Zhizheng Zheng,
        Rulin Xu, 
        Yitong Liu, 
        Si Zheng, 
        Xiangke Liao
    }
    \IEEEauthorblockA{
        \textit{College of Computer Science and Technology}\\
        \textit{National University of Defense Technology}\\
        Changsha, China \\
        \{zhangyuanliang13, xieyifan, shanshanli, liuke23, jiazhouyang, xbhuang, songj19, luochaopeng18, zhengzhizheng23, \\xurulin11, liuyitong22, xkliao\}@nudt.edu.cn, \{ridicious1997, si.zheng1009\}@gmail.com
    }
    \thanks{\IEEEauthorrefmark{1}Co-first authors}
    \thanks{\IEEEauthorrefmark{2}Corresponding author}
}

\maketitle
\begin{abstract}

Recently, large language models (LLMs) have shown strong potential in code generation tasks. However, there are still gaps before they can be fully applied in actual software development processes. Accurately assessing the code generation capabilities of large language models has become an important basis for evaluating and improving the models. Some existing works have constructed datasets to evaluate the capabilities of these models. However, the current evaluation process may encounter the illusion of ``Specialist in Familiarity'', primarily due to three gaps: the exposure of target code, case timeliness, and dependency availability. The fundamental reason for these gaps is that the code in current datasets may have been extensively exposed and exercised during the training phase, and due to the continuous training and development of LLM, their timeliness has been severely compromised.

The key to solve the problem is to, as much as possible, evaluate the LLMs using code that they have not encountered before. Thus, the fundamental idea in this paper is to draw on the concept of code obfuscation, changing code at different levels while ensuring the functionality and output. To this end, we build a code-obfuscation based benchmark \datasetO{}. We first collect \checked{1,354} raw cases from \checked{five} real-world projects, including function description and code. Then we use three-level strategy (symbol, structure and semantic) to obfuscate descriptions, code and context dependencies. We evaluate \checked{four} LLMs on \datasetO{} and compared the effectiveness of different obfuscation strategy. We use official test suites of these projects to evaluate the generated code. The results show that after obfuscation, the average decrease ratio of test pass rate can up to \checked{62.5\%}. 

\end{abstract}

\begin{IEEEkeywords}
Large Language Model, Code Generation Capability, Code Dataset
\end{IEEEkeywords}

\section{Introduction}

\label{sec:intro}

With the rapid development of the Large Language Model (LLM), the code generation capability of LLMs has attracted lots of attention~\cite{liu2024empirical,wang2023review,lin2024llm,kazemitabaar2023novices,koziolek2024llm,liu2023your,xia2024top,liu2024evaluating}. However, how to accurately assess the code generation capability of LLMs in production-level software development is still an open question. A variety of benchmark tests for code generation have been proposed, however, there are still gaps between these benchmark tests and the actual software development process. 


Traditional datasets as evaluation benchmarks~\cite{chen2021evaluating,cassano2022multipl,hao2022aixbench,lai2023ds} play key roles in evaluating the capabilities of LLMs. However, they focus mainly on standalone functions based on algorithmic problems, which cannot reflect the complexity of real software development, as LLMs may face potential challenges in handling code that is interdependent with other contextual elements of the project~\cite{yu2024codereval}. There have been several benchmarks~\cite{yu2024codereval,jimenez2023swe, li2024deveval,zhu2024domaineval} which are built based on real-world production projects, while there remain three gaps for these benchmarks to objectively evaluate the code generation capability of LLM: exposure of target code, case timeliness, and dependency availability.

\begin{figure}
	\centering
    \vspace{-1mm}
	\includegraphics[width=0.45\textwidth]{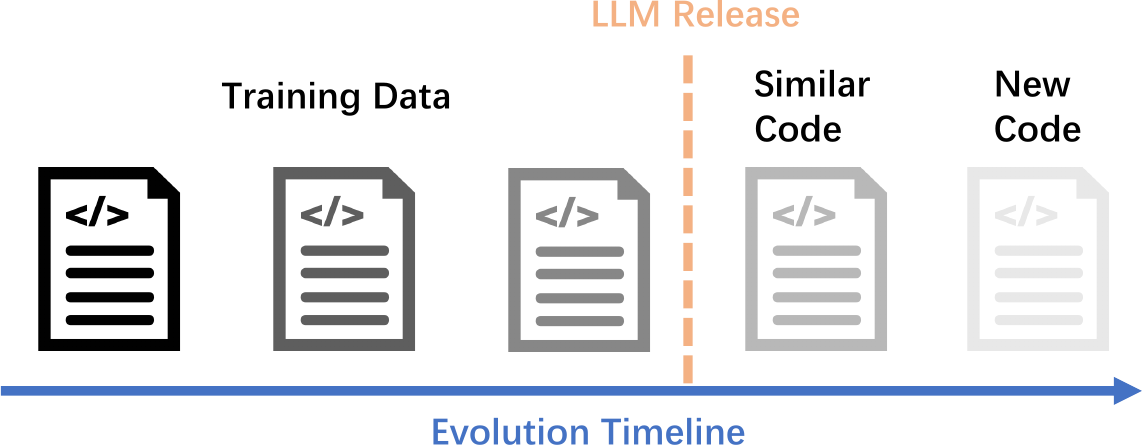}
	\caption{The familiarity level of code to LLM. The more transparent elements indicate less familiarity.}
    \vspace{-1mm}
	\label{fig:intro}
  \vspace{-2mm}
\end{figure}

\noindent\textbf{Gap 1: Target code has been exposed in the pre-training stage.}
Previous benchmarks~\cite{yu2024codereval, li2024deveval, zhu2024domaineval} have only rewritten the functional descriptions without modifying the code itself. This could result in the target code being exposed to LLMs during training, making the evaluation results not objective as the code has been exercised extensively. 

\noindent\textbf{Gap 2: Collected case is time-sensitive.} With the rapid development of LLMs, data will be continuously trained. Although previous benchmark~\cite{jimenez2023swe} has collected modified code from history after LLMs training, the modified code may still have been exposed to LLMs because of the existence of code clones. In addition, benchmarks that rely on historical data will suffer from the timeliness problem and cannot be used to evaluate subsequent releases of LLMs.

\noindent\textbf{Gap 3: Precise dependencies are difficult to provide in real usage scenarios.}  
Existing benchmarks~\cite{li2024deveval,yu2024codereval} directly provide the model with all the dependencies needed to generate the target code. However, such conditions can not be always satisfied in real development.

\textbf{In conclusion, existing evaluation process may suffer from the ``Specialist in Familiarity'' problem}, which means LLMs can perform well on code that they are very familiar with, often due to extensive experience or repeated exposure. This term highlights the individual's strength in specific, well-known domains, but it also implies that their expertise may not extend beyond these familiar areas. The primary reason is that the code used to evaluate LLMs may be extensively exposed and exercised for training. As shown in \figurename~\ref{fig:intro}, the unfamiliarity of code to LLM is increased along with the timeline. Unfortunately, even for the code collected after LLM training and release may be similar to the training data due to the existence of code reuse. 

To solve these problems, we need to evaluate using code that LLMs have not encountered before as much as possible. To this end, we propose a new obfuscation-based benchmark \datasetO{} to evaluate LLMs' code generation capability. Our main target is to evaluate the LLM on the code generation tasks that it has never encountered before and simulate real software development process.
\datasetO{} has three key characteristics: 1) \textbf{To solve Gap 1}, we collect highly starred projects from GitHub \cite{github} and select functions from them that are introduced after a certain time point. These functions are covered by the official test suites. 2) \textbf{To solve Gap 2}, all the raw data have been obfuscated by different-level strategies (symbol, structure, and semantic) to rewrite both functional descriptions and code. The process of code obfuscation can ensure that datasets are reused without the concern that they might still become training data in the future (the code obfuscation process can be repeated). 3) \textbf{To solve Gap 3}, we provide relevant code dependencies in a compromise manner, simulating real-world development scenarios without deliberately sacrificing the generation capabilities of LLM. We identify all the contextual dependencies necessary for each function, including necessary API calls, structures, macros, etc., and also add some dependencies that are not related to the target function code for obfuscation.

To effectively utilize our dataset and evaluate LLM's real capability on generating unfamiliar code, we built a project-level execution platform that provides an off-the-shelf runtime environment to automatically evaluate the functional correctness of the generated code. We developed this platform based on Docker, cloning and building the environment for all projects. Given a model-generated code, the code will automatically replace the original code. Then the projects will be compiled and tested to see whether there are compilation errors or test failures. 

We comprehensively evaluated four state-of-the-art code generation models (ChatGPT3.5, ChatGPT4-1106, ChatGPT4-0125, and DeepSeek-Coder-V2) on \datasetO{}. We analyzed each model's effectiveness under different obfuscation strategies. The results show that after code obfuscation, the average decrease ratio of test pass rate can up to \checked{62.5\%}.
In addition, we found that even passing all the tests, code generated by LLMs may still suffer from non-functional code issues (e.g., code robustness), which can guide the developers to better discern and utilize the code generated by LLMs.

The main contributions of the paper are as follows:

\begin{itemize}[leftmargin=*, topsep=1pt, itemsep=1pt]

  \item  We reveal that existing benchmarks are insufficient for objectively evaluating the code generation capabilities of LLMs, primarily due to three gaps: exposure of target code, case timeliness, and dependency availability.

 \item We propose an obfuscation-based approach to rewrite the functional descriptions, code, and dependencies to prevent the target code from being exposed in the training stage. We design different levels of obfuscation strategies and examine their effectiveness. Future research can design sophisticated obfuscation process to better explore the potential of LLM's capability based on our results.

 \item We build an obfuscation-based benchmark \datasetO{}\footnote{https://github.com/zhangbuzhang/ObfusEval} using code from real-world projects. We evaluated four state-of-the-art code generation models on \datasetO{}. The results show that after code obfuscation, the average decrease ratio of test pass rate is \checked{15.3\%-62.5\%}, demonstrating the inflated capabilities of LLMs. We also identify non-functional code issues in the passed cases which can be studied in future work.

\end{itemize}

\section{BACKGROUND}
\label{sec:related}

In this section, we first conduct a comprehensive examination of the latest advancements in Large Language Models (LLMs) within code generation. Subsequently, we delve into the related work of evaluations crafted for code generation, along with the limitations and challenges encountered in assessing the performance of LLMs. Finally, we introduce the motivation for incorporating code obfuscation in the evaluation of large language models.

\subsection{Large Language Models for Code Generation.}

The process of code generation, which involves the automatic creation of complete program code or the completion of code snippets from higher-level representations, such as natural language descriptions, models, or specifications, plays a pivotal role in enhancing programming efficiency and mitigating human error~\cite{budinsky1996automatic,svyatkovskiy2020intellicode,huang2023towards,geng2024large}. 
Recent advancements in Large Language Models (LLMs) for code generation have garnered significant attention in the realm of computer science research. These LLMs, such as GPT-4~\cite{GPT-4}, ChatGLM~\cite{du2021glm}, CODEX~\cite{chen2021evaluating}, and CodeGen~\cite{nijkamp2022codegen}, have demonstrated remarkable capabilities not only in general natural language processing tasks~\cite{chang2023survey} but also in the specific area of code generation. Notably, GPT-4 achieved the highest pass rate on the HumanEval benchmark~\cite{chen2021evaluating}, indicating a growing trend to evaluate the code generation capacity of general LLMs~\cite{shen2023pangu}.

Code-specific LLMs, which are trained primarily on massive code-specific corpora, often outperform general LLMs in code generation tasks~\cite{lozhkov2024starcoder,zan2023large,luo2023wizardcoder,shen2023pangu,guo2024deepseek,ma2023training,lin2023cct5,dong2023codep}. Diverse training approaches have been employed, with some models like InCoder~\cite{fried2022incoder} and StarCoder~\cite{li2023starcoder} being trained with the "filling-in-the-middle" capability for infilling missing code based on context. Varieties of code LLMs have been proposed, such as WizardCoder~\cite{luo2023wizardcoder}, Instruct-StarCoder~\cite{Instruct-starcoder}, and Instruct-CodeGen~\cite{Instruct-codegen}, each designed with different training objectives.

\begin{table}[t]
\renewcommand\arraystretch{1.5}
\caption{LLM's code generation performance on code competition problems of different time}
\begin{center}
\setlength\tabcolsep{4pt}
\begin{tabular}{ccccc}
\hline
\multirow{2}*{\textbf{OJ Website}} & \multicolumn{2}{c}{\textbf{GPT3.5-turbo (2023.06)}} & \multicolumn{2}{c}{\textbf{GPT4.0-preview (2023.11)}}\\
\cline{2-5}
~ & \textbf{zero-shot} & \textbf{few-shot} & \textbf{zero-shot} & \textbf{few-shot} \\
\hline
\textbf{LeetCode} & \multirow{2}{*}{19.7\%} & \multirow{2}{*}{22.0\%} & \multirow{2}{*}{39.4\%} & \multirow{2}{*}{40.9\%}\\
\textbf{2023.11-2024.01} \\ 
\hline
\textbf{LeetCode} & \multirow{2}{*}{95.6\%}& \multirow{2}{*}{93.3\%} & \multirow{2}{*}{96.7\%} & \multirow{2}{*}{95.6\%} \\
\textbf{2018.09-2018.11} \\
\hline
\end{tabular}
\end{center}
\label{tab:OJresult}
\vspace{-3mm}
\end{table}

\subsection{Evaluations for LLM's Code Generation}

\textbf{Benchmark construction. }Current benchmarks built based on real projects usually rewrite only the functional descriptions without modifying the code~\cite{li2023skcoder,yu2024codereval,jimenez2023swe,li2024deveval,hao2022aixbench,huang2024karecoder}, which may lead to the ``code leak'' issue. SWE-BENCH~\cite{jimenez2023swe} collects modified code from the project's history. However, the code may be in the training set of subsequent releases of LLMs. Future LLMs may become experts in solving problems in SWE-BENCH, but they may still struggle with new code problems in real-world scenarios. EvoCodeBench~\cite{li2024evocodebench} periodically update the dataset, but there is still possibility of introducing similar code due to the existence of code clones.

\textbf{Context dependencies. }
Traditional benchmarks focus on generating independent code units, ignoring the contextual relationships between code~\cite{chen2021evaluating,austin2021program,zheng2023codegeex,li2022competition,yu2018spider, he2024beyond}. For example, SWE-BENCH does not provide the complete dependencies of the generated code, so it is hard to distinguish the capability of LLMs to generate the target code and its dependencies. However, current studies indicate that only about 30\% of methods in open-source projects are relatively independent~\cite{yu2024codereval} in real-world scenarios, methods often depend on each other or share variables, which is not considered in these traditional benchmarks. Some works~\cite{yu2024codereval,li2024deveval} provide complete dependencies but do not include useless and obfuscated dependencies. This discrepancy does not align with software development scenarios and fails to accurately measure the ability of LLMs to assist developers in practical settings.
Therefore, we need evaluation methods closer to real-world scenarios to comprehensively assess LLMs' performance in real-world software development.

\textbf{Evaluation methods.} When evaluating the code generation capabilities of LLMs, existing studies and benchmarks focus on the basic correctness of the code, which is usually verified by executing simple test cases (e.g., unit tests)\cite{liu2024exploring, spiess2024calibration,wu2024benchmarking}. In our work, we try to assess both syntactic and functional correctness of LLMs-generated code in real scenarios, by leveraging both compile checking and systematic testing, which aligns more closely with the requirements of real-world development. 
In addition, previous benchmarks were compared to Humaneval\cite{chen2021evaluating} to prove their validity \cite{du2023classeval,li2024deveval,yu2024codereval,li2024acecoder}. However, since the length of code to be generated by these benchmark tests does not match the distribution of code lengths in Humaneval, this comparison is inherently unfair. It therefore does not accurately reflect the validity of the benchmark tests.

\subsection{Motivation of Using Code Obfuscation}

As LLMs are continuously trained and released, traditional datasets would constantly be learned and trained by these models. Therefore, the timeliness of code may be a crucial factor when testing the generation capacity of LLM. In other words, code may become easier to generate because it already exists (or similar) in the training set.

To validate our conjecture, we conduct a pilot study to check whether the timeliness of code will affect the result of LLM generation capability. We use GPT-3.5-turbo (released at 2023.06.13) and GPT 4.0-preview (released at 2023.11.06) to do code competition problems from LeetCode \cite{leetcode}. We collect the problems from 2018.09 to 2018.11 (90 problems) and the problems from 2023.11 to 2024.01 (127 problems) as our test data. The later 127 problems came out after the release of two models, which were theoretically not presented in the model's training data.

Table~\ref{tab:OJresult} shows the results. We find that the pass rates of the early 90 problems are much higher than later problems. Due to the fact that code competition websites contain problems of varying difficulty levels, and there is no deliberate increase of the difficulty of new problems, the reason is that these older problems appeared in the model's training dataset so that models can handle them more easily.

\begin{figure}
    \centerline{\includegraphics[width=0.5\textwidth]{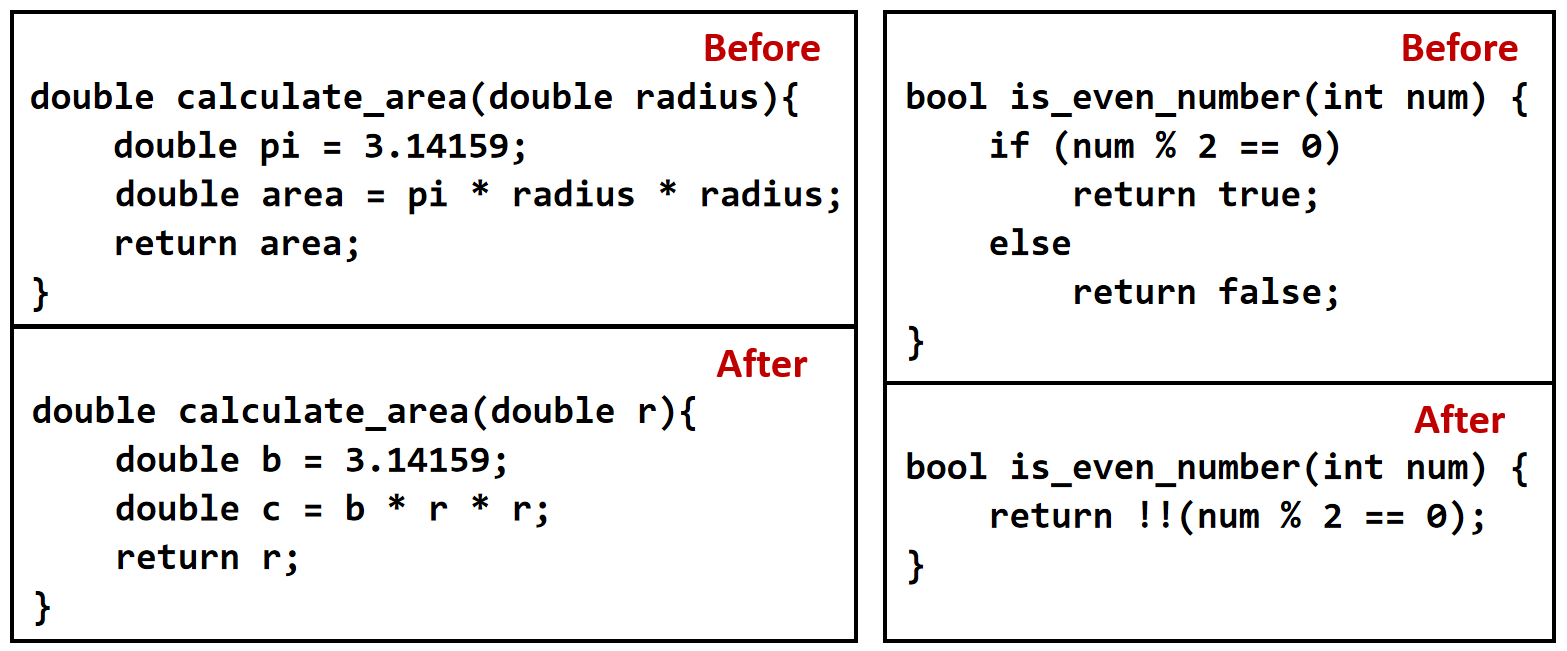}}
    \caption{Examples of code obfuscation.}
    \label{fig: code obfuscation}
     \vspace{-2mm}
\end{figure}

Relying solely on the latest code as a dataset is not a sustainable approach in the long run (as models will continuously train and evolve), and due to the existence of code reuse, even code written after the training cutoff date may still be similar to code in the training set. During the construction of our dataset, to mitigate the impact that LLMs may have seen the code, we have drawn inspiration from code obfuscation techniques~\cite{schrittwieser2011code,schrittwieser2016protecting,balakrishnan2005code}, which are originally used for making applications difficult to be decompiled or disassembled. \figurename~\ref{fig: code obfuscation} illustrates two examples of code obfuscation (changing the variable names and changing the code implementation of the same logic). After obfuscation, the code will become different from the training set while maintaining its functionality and output. 

\section{Benchmark Construction and Testing}
\label{sec:research}

\begin{figure*}[t]
	\centering
    \vspace{-1mm}
	\includegraphics[width=1\textwidth]{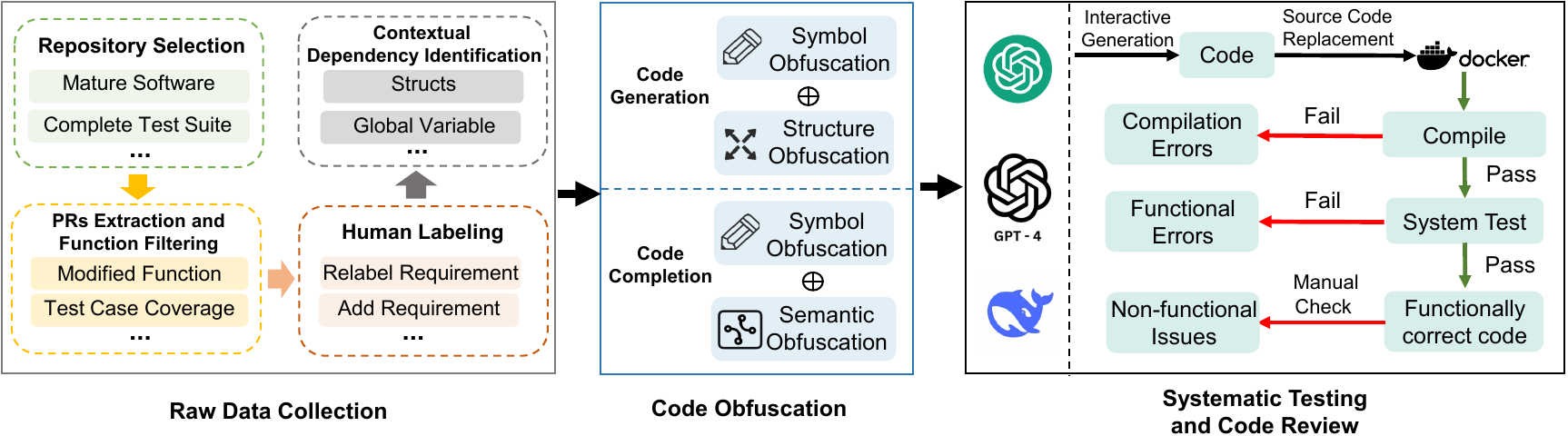}
 \vspace{-4mm}
	\caption{Workflow of dataset construction and testing}
	\label{fig:overview}

\end{figure*}

In this section, we describe the approach of building and testing \datasetO{} to show the capability of LLM, and the workflow is shown in ~\figurename~\ref{fig:overview}.
The building process mainly includes two phases: 
collecting raw data from open-source projects (Section~\ref{Sec4.1}), and obfuscating code (Section~\ref{Sec4.2}).
After that, we construct a testing framework to evaluate the effectiveness of code generation and code completion by the models on \datasetO{} (Section~\ref{Sec4.3}).

\subsection{Raw Data Collection}\label{Sec4.1}

To construct \datasetO{}, we first need to collect the raw data. The raw data includes functions modified after the large model training data cutoff date from real-world open source projects, as well as the context information these functions depend on within the project. 
Overall, we divide the raw data collection process into two parts: function collection and contextual dependency provision.

\subsubsection{Functions Collection}\label{Sec4.1.1}

Function collection includes three main steps:

\textbf{\textit{Step 1: Repository Selection.}}
The existing dataset mainly covers Java and Python projects. 
To demonstrate the code generation capabilities of the large language models in other programming languages, we chose C projects to build \datasetO{}.
We set two conditions to filter repositories from GitHub: 1) The repository should be mature and well-maintained; 2) The project should have a comprehensive test suite for systematic testing.
Finally, we selected five projects~\cite{redis,libgit2,libvips,fluent,lvgl}, with over 20k average stars.

\textbf{\textit{Step 2: PRs Extraction and Function Filtering.}}
We selected merged PRs from the chosen repositories and extract functions that meet the task criteria. 
Specifically, we extracted PRs and filter functions based on the following four criteria:
1) The PRs are merged after a certain time point to coincide with the training data cutoff date of specific models we tested; (more discussion in Section~\ref{sec:data});
2) The PRs modify the repository's test files to verify the modified code's functional correctness;
3) The functions are modified in the PR to ensure that LLM had not seen modified code in previous code base;
4) The functions are covered by the test suites to ensure that the functionality of the functions is effectively verified.
\Space{
\begin{itemize}[leftmargin=*, topsep=1pt, itemsep=1pt]
  \item \textbf{Criteria 1:} We select PRs with merge dates after a certain time point to coincide with the training data cutoff date of specific models we tested. (more discussion in section~\ref{sec:data})
  \item \textbf{Criteria 2:} We select functions modified in the PR to ensure that large models had not seen modified code in previous code base.
  \item \textbf{Criteria 3:} We ensure that the selected PRs modify the repository's test files to verify the modified code's functional correctness.
  \item \textbf{Criteria 4:} We select the functions covered by the test suite to ensure that the functionality of the functions is effectively verified.
\end{itemize}
}

Based on the above criteria, we finally selected 1,354 functions for constructing code generation and code completion tasks and the statistical results are shown in Table~\ref{tab:selectProjectStatics}.

\begin{table}[t]
  \centering
  \caption{The statics of PRs Extraction and Function Filtering.}
   \vspace{-1mm}
    \begin{tabularx}{0.48\textwidth}{Xcccc}
    \toprule
    \multirow{2}{*}{\textbf{Software}} & \textbf{Merged} & \textbf{PRs with} & \textbf{Modified} & \textbf{Test-Covered} \\
    & \textbf{PRs} & \textbf{Tests Modified} & \textbf{Functions} & \textbf{Functions} \\
    \midrule
    redis   & 740   & 114   & 3,142  & 681 \\
    libvips & 236   & 14    & 1,285  & 203 \\
    lvgl    & 1,718  & 36    & 1,447  & 303 \\
    libgit2 & 118   & 17    & 618   & 78 \\
    fluent  & 160   & 46    & 419   & 89 \\
    \midrule
    Total   & 2,972  & 227   & 6,911  & 1,354 \\
    \bottomrule
    \end{tabularx}
  \label{tab:selectProjectStatics}

\end{table}

\textbf{\textit{Step 3: Human Labeling.}}
In this step, we manually provide functional descriptions for each test-covered function.
Specifically, we assembled a team of seven senior software engineers, each with at least five years of C programming experience.
The team is responsible for rewriting the existing functional descriptions and providing manually written descriptions for functions without descriptions, aiming to reduce the model's dependence on the original functional descriptions encountered during the pre-training phase.
During this process, we implemented a double-check mechanism.
When any two engineers have a disagreement, a third engineer is brought in to discuss and reach a final consensus together.

\subsubsection{Contextual Dependency Provision}

Previous work~\cite{yu2024codereval} has shown that more than 70\% of the functions depend on other contextual information in the project, therefore, the inability to provide dependencies can lead to a significant decline in the generative capabilities of large language models. However, accurately providing the dependencies required for code generation is extremely difficult and does not align with real-world development scenarios. Consequently, we adopt a conservative approach to providing dependencies.

We first use syntax tree analysis to identify and collect all relevant contextual dependencies in the code. we extract dependencies from project files, including the names of functions, declarations, function bodies, global variables, structures, macros, as well as function comments. Next, we compile the code to obtain the LLVM IR intermediate code representation of the files. By matching keywords in the IR syntax (such as "call" to indicate a function invocation), we traverse the IR files to acquire the names of those dependencies. By cross-referencing the results from the first step, we can obtain the contextual information of the target function. We also provide similar but different dependencies for each contextual dependency to simulate the disturbances caused by irrelevant information in the actual development process.

\begin{table*}[t]
  \centering
  \caption{The composition of \datasetO{}}
   \vspace{-1mm}
    \begin{tabularx}{1\textwidth}{Xcccccc}
    \toprule
    \multirow{2}{*}{\textbf{Soft.}} 
    & \textbf{Original} 
    & \textbf{Symbol Obfuscation} 
    & \textbf{Structure Obfuscation} 
    & \textbf{Semantic Obfuscation} 
    & \textbf{Symbol + Structure}
    & \textbf{Symbol + Semantic} \\
    &\textbf{Functions}
    & \textbf{Functions} 
    & \textbf{Functions} 
    & \textbf{Functions} 
    & \textbf{Obfuscation Functions}
    & \textbf{Obfuscation Functions}\\
    \midrule
    redis   & 681 & 681 & 215 & 106 & 215 & 106 \\
    libvips & 203 & 203 & 58  & 17  & 58  & 17  \\
    lvgl    & 303 & 303 & 115 & 15 & 115  & 15  \\
    libgit2 & 78  & 78  & 32  & 10  & 32  & 10  \\
    fluent  & 89  & 89  & 30  & 11  & 30  & 11  \\
    \midrule
    Total   & 1,354 & 1,354 & 450 & 159 & 450 & 159 \\
    \bottomrule
    \end{tabularx}
  \label{tab:statics}
   \vspace{-2mm}
\end{table*}

\subsection{Code Obfuscation}\label{Sec4.2}

\begin{figure*}[htbp]
	\centering
    \vspace{-1mm}
	\includegraphics[width=1\textwidth]{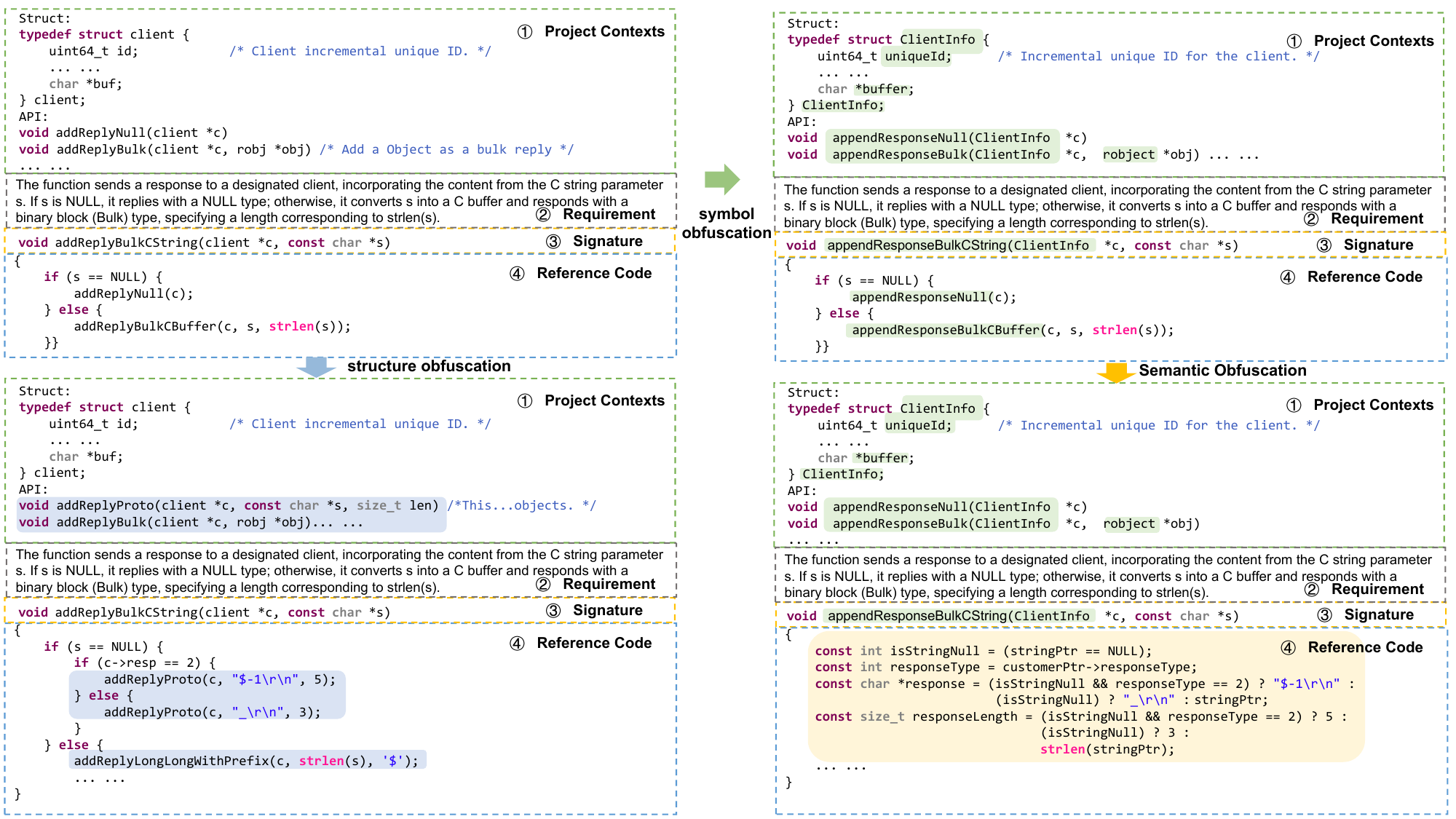}
	\caption{An example of code obfuscation process}
    \vspace{-1mm}
	\label{fig:confuseProcess}
 \vspace{-2mm}
\end{figure*}

Despite selecting code from the project revision history that was modified after the training time of the large model, we also applied additional code Obfuscation techniques to the dataset to enhance protection against "code leakage" and ensure the applicability of our benchmarks in future releases of the large model.
\figurename~\ref{fig:confuseProcess} shows the example process of Obfuscating the dataset with three strategies. 
To objectively evaluate the effectiveness of the LLM when dealing with obfuscated code, we constructed code generation and code completion scenarios, considering the code completion task as a subtask of the code generation task.
We apply symbol obfuscation and structure obfuscation strategies in code generation scenarios and symbol obfuscation and semantic obfuscation strategies in code completion scenarios (the strategies can be used in combination).

Distinguishing different obfuscation strategies for various task scenarios is due to the fact that only in the code completion scenario can semantically obfuscated code fragments be retained and a LLM be required to generate complete code. This is to test the LLM's true generation capability when faced with semantically obfuscated code. 
We do not fully integrate the three types of obfuscation together because different obfuscation strategies are suitable for different types of code. Next, we will discuss in detail the effects and practice of each obfuscation.

\subsubsection{\textbf{Symbol Obfuscation}}
 
We first use a comprehensive symbol obfuscation strategy to not only rewrite the functional descriptions of the functions but also perform thorough identifier rewrites for all meaningful identifiers in the target code and the provided context.
This means that all identifiers for functions, variables, class names, etc. in the code are obfuscated regardless of the context in which they appear in the source code. We use NLTK~\cite{nltk} to do the word segmentation and replacement. This strategy is designed at the token-level to change the LLM's familiarity to the target code. 

\subsubsection{\textbf{Structure Obfuscation}}

After symbol obfuscation, we further change the code structure automatically using a structure obfuscation strategy. In this stage, we employ the strategy to adjust and integrate the calling structure of the target function so that the execution path and organization of the function are changed. Specifically, we use LLVM~\cite{llvm} to unfold and integrate the functions that are called in the objective function to change the structure of the code. We extract all the called functions and their implementations within the target function based on the context, and then utilize the abstract syntax tree (AST) to handle parameter passing and automatically unfold the called functions.

\subsubsection{\textbf{Semantic Obfuscation}}

In semantic obfuscation, we meticulously rewrite code snippets within functions to ensure that the new code is semantically equivalent to the original, yet the implementation logic differs. The goal of this process is to maintain the functional consistency while introducing a new implementation method, effectively obfuscating the code at the method-level. Through semantic transformation, we ensure that the code can still achieve the same functionality, but for the model, its generative logic is completely different from the original code. This semantic obfuscation provides a more challenging task to test the code understanding and generative capabilities of LLM. Since this obfuscation method relies on specific semantics of the code, the current approach involves manually rewriting the code and conducting a double-check.
We highly recommend that future research should delve deeper into semantic obfuscation strategies and design templates to automate this process.

Through raw data collection and code obfuscation, we constructed the \datasetO{}, which is shown in Table~\ref{tab:statics}. Apart from symbol obfuscation, we did not apply structure and semantic obfuscation to all the original data. This is partly because the characteristics of the code may be suitable for specific obfuscation methods (such as nested structures), and partly due to the cost of manual inspection. The code examples from Redis are more numerous and regular, so we applied more semantic obfuscation to them. Future research could design automated code obfuscation framework for obfuscation and inspection.

\subsection{Systematic Testing and Code Review}\label{Sec4.3}

To evaluate the LLM's code generation capabilities on the \datasetO{} and the effectiveness of our code obfuscation methods, we designed an automated code execution and verification platform. 
The platform is built on Docker images, providing an isolated sandbox environment to ensure that the tested codes do not interfere with each other.
The evaluation process of generated code mainly includes two parts: systematic testing and code review.

\subsubsection{Systematic Testing}

We use the systematic testing process to evaluate the model and the dataset. We utilize compilation checks and official test suites to detect syntax and semantic errors. 
We construct the prompt to guide the large model in code generation scenarios and code completion scenarios.
The composition of the prompt is detailed below:

\begin{itemize}[leftmargin=*, topsep=1pt, itemsep=1pt]
    \item \textbf{Instruction:} We provide explicit instructions to guide the LLMs to generate code related to the software.
    \item \textbf{Context}: We provide detailed context information, including structs, macros, functions, global variables, etc. 
    These contexts include both the necessary dependencies for implementing the target function and the context that is irrelevant to the target function.

    \item \textbf{Function description}: We provide a functional description of the target function to guide LLMs in generating code. Based on this description, we provide only the declaration of the target function for \textbf{code generation scenarios} and partial code implementation details of the target function for \textbf{code completion scenarios}.
\end{itemize}

\begin{figure}[t]
	\centering
        \vspace{-1mm}
	\includegraphics[width=0.5\textwidth]{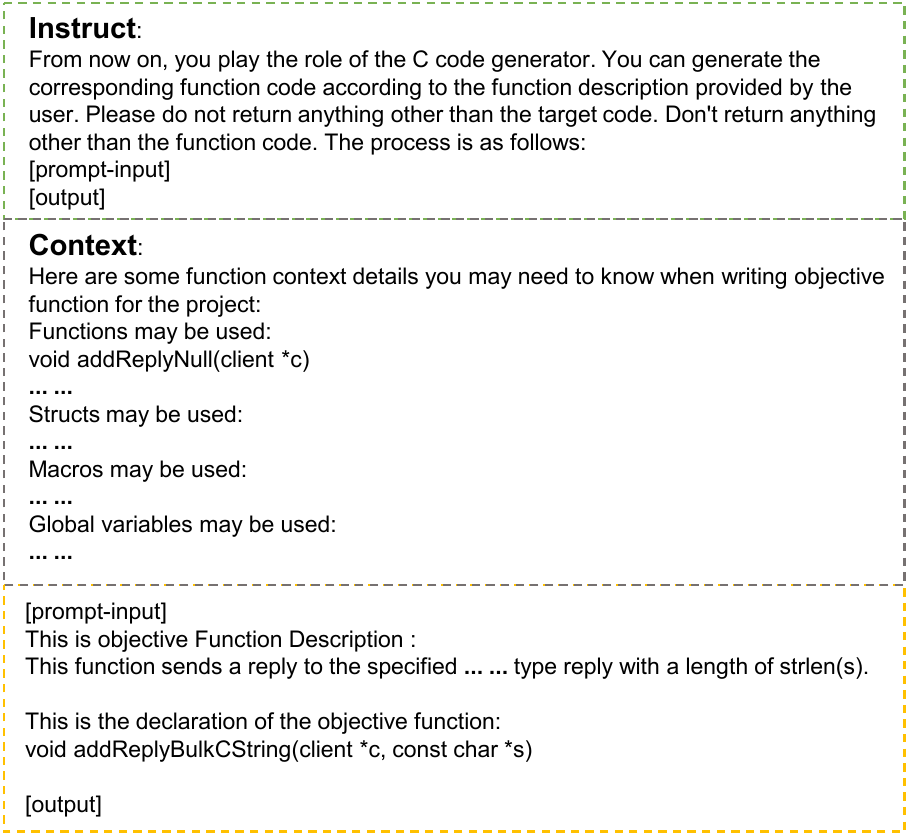}
	\caption{Example of prompt for code generation scenarios}
\vspace{-4mm}
	\label{fig:codeGen} 
\end{figure}

Fig.~\ref{fig:codeGen} shows an example of the prompt for the code generation scenario. 
We provide functional descriptions in the [prompt input] to guide the large model's code generation.
After obtaining the code generated by the model, we integrate it into the software for compilation and system testing. 
If errors occur, we separately record compilation errors and test errors, then analyze the error information. For code that passes the tests, we will conduct a manual code review.

 \subsubsection{Code Review}

We used compilation checks and test suites to evaluate the syntactic correctness and functional correctness of the generated code, respectively. However, through manual inspection, we found that even the code was functionally correct, i.e., it was able to pass the tests, there were still some non-functional code quality issues. These problems may include deficiencies in code efficiency, code robustness, etc. Therefore, we manually analyzed these non-functional code quality issues (Section~\ref{sec:common issues}).

\section{Evaluation}\label{sec:setup}

\begin{table*}[t]
  \centering
  \caption{\textbf{The performance of LLMs in code generation scenarios.} "Original" means the pass rate of all raw code tasks. "Original(Structure)" means the pass rate of those raw code tasks which later perform structure obfuscation. All the numbers have omitted the percentage sign.}
  \resizebox{0.9999\linewidth}{!}{
    \begin{threeparttable}
    \begin{tabular}{cccccc|cccccc}
    \toprule
    \multirow{2}[4]{*}{\textbf{Software}} & \multirow{2}[4]{*}{\textbf{Model}} & \multicolumn{2}{c}{\textbf{Original}} & \multicolumn{2}{c|}{\textbf{Symbol}}  & \multicolumn{2}{c}{\textbf{Original (Structure)}} & \multicolumn{2}{c}{\textbf{Structure}} & \multicolumn{2}{c}{\textbf{Symbol+Structure}} \\
\cmidrule{3-12}          &       & \textbf{CPR} & \textbf{TPR} & \textbf{CPR} & \textbf{TPR} & \textbf{CPR} & \textbf{TPR}& \textbf{CPR} & \textbf{TPR} & \textbf{CPR} & \textbf{TPR} \\
    \midrule
    \multirow{5}[2]{*}{redis} 
          & GPT3.5 &38.2 & 11.9  &19.0~\dec{}& 9.7~\sdec{}  &44.7&  10.7  &36.3~\sdec{}& 6.5~\dec{}&29.3~\dec{}&  2.9~\dec{}\\
          & GPT4-1106 &37.0 & 17.6  &31.7~\sdec{}& 17.1~\sdec{}  &34.4& 13.9  & 36.5~\sinc{} & 7.0~\dec{}&19.6~\dec{}& 4.7~\dec{}  \\
          & GPT4-0125 &39.9 & 20.4  &31.8~\sdec{}& 17.3~\sdec{}&53.0&13.9  &34.9~\dec{}&9.8~\dec{}&19.5~\dec{}&  2.8~\dec{} \\
          & DeepSeek & 39.6 & 7.2  & 28.8~\sdec{}& 11.5~\inc{}&47.4 &5.1 & 32.1~\dec{} & 5.1~\equa{} &22.4~\dec{}&4.7~\sdec{} \\
        \cmidrule{2-12}    
          & \textbf{Average} & 38.7 & 14.3 & 27.8~\sdec{28.2\%} & 13.9~\sdec{2.8\%}&44.9 &10.9& 35.0~\sdec{22.0\%} &7.1~\dec{34.9\%}&22.7~\dec{49.4\%}&3.8~\dec{65.1\%}\\
    \midrule
    \multirow{5}[2]{*}{libvips} 
          & GPT3.5 &58.6&  18.2   &44.8~\sdec{}&  18.7~\sinc{} &70.7& 20.7  &74.1~\sinc{}&  20.7~\equa{}&46.6~\dec{}&12.1~\dec{} \\
          & GPT4-1106 &42.9 & 21.7  & 32.5~\sdec{}  & 19.2~\sdec{}&44.8 & 15.5  & 44.8~\equa{}& 20.7~\inc{} &27.6~\dec{}&13.8~\sdec{}\\
          & GPT4-0125 & 43.8  & 25.6  & 41.4~\sdec{} & 22.2~\sdec{}&51.7 & 27.6  & 50.0~\sdec{} & 24.1~\sdec{}  &48.3~\sdec{} &19.4~\sdec{} \\
          & DeepSeek &50.8& 29.6   &41.3~\sdec{}&  24.6~\sdec{} &53.5& 25.9  &56.9~\sinc{}&  22.4~\sdec{}&44.8~\sdec{}& 15.5~\dec{} \\
          \cmidrule{2-12}
          & \textbf{Average} &49.0&23.8  &40.0~\sdec{18.4\%}&21.2~\sdec{10.9\%}&55.2& 22.4  &56.5~\sinc{2.4\%}& 22.0~\sdec{1.8\%}  &41.8~\sdec{24.3\%}&15.2~\dec{32.1\%}\\
    \midrule
    \multirow{5}[2]{*}{lvgl} 
          & GPT3.5 &  36.7 &19.8  & 35.8~\sdec{}  & 17.5~\sdec{} &27.3&11.6  &27.0~\sdec{} &6.9~\dec{}  &13.9~\dec{}&6.9~\dec{}  \\
          & GPT4-1106 &38.9 &26.7 & 39.7~\sinc{} &24.8~\sdec{} &22.6 & 13.9 & 23.5~\sinc{}  &12.1~\sdec{} &16.5~\sdec{} &10.4~\sdec{}  \\
          & GPT4-0125 &46.2 &31.0  & 43.3~\sdec{}&28.1~\sdec{} &28.7  & 15.7 & 27.0~\sdec{} &11.3~\sdec{}& 20.0~\dec{}&11.3~\sdec{} \\
          & DeepSeek &44.2 &30.4 & 42.6~\sdec{}  &28.4~\sdec{} &29.6 & 14.8 & 28.7~\sdec{} & 9.6~\dec{}  & 17.4~\dec{} & 7.0~\dec{}  \\
          \cmidrule{2-12}
          & \textbf{Average} & 41.5 & 27.0   &40.4~\sdec{2.7\%} &24.7~\sdec{8.5\%}&27.1&14.0 &26.6~\sdec{1.8\%} & 10.0~\sdec{28.6\%} &17.0~\dec{37.3\%}  &   8.9~\dec{36.4\%}\\
    \midrule
    \multirow{5}[2]{*}{libgits}
          & GPT3.5 &14.1& 14.1  &11.5~\sdec{}& 7.7~\dec{}  &13.1& 13.1   &12.5~\sdec{}& 12.5~\sdec{}  &6.2~\dec{}&  3.1~\dec{}     \\
          & GPT4-1106 &38.5& 23.1  &18.0~\dec{}& 9.0~\dec{}  &31.3& 25.0  &12.5~\dec{}&  12.5~\dec{}  &0.0~\dec{}& 0.0~\dec{}  \\
          & GPT4-0125 &39.7& 20.5  &12.8~\dec{}& 6.4~\dec{}  &34.4& 18.8  &12.5~\dec{}&  12.5~\dec{}  &0.0~\dec{}& 0.0~\dec{}   \\
          & DeepSeek &23.1&  21.8   &14.1~\dec{}& 12.8~\dec{} &18.8&15.6   &15.6~\sdec{}&  15.6~\equa{}     &18.6~\sdec{}& 9.2~\dec{}   \\  
          \cmidrule{2-12}
          & \textbf{Average} &28.9&19.9&14.1~\dec{51.2\%}&9.0~\dec{54.8\%}&24.4&18.1 &13.3~\dec{45.5\%}&13.3~\sdec{26.5\%}&6.2~\dec{74.6\%}&3.1~\dec{82.9\%}\\
    \midrule
    \multirow{5}[2]{*}{fluent} 
          & GPT3.5 &27.0& 19.1  &18.0~\dec{}&  9.0~\dec{}  &33.3&30.0  &23.3~\dec{}& 10.0~\dec{}  &30~\sdec{}& 3.3~\dec{} \\
          & GPT4-1106 &25.3& 20.2  &12.4~\dec{}& 10.1~\dec{}  &30.0&26.7   &13.3~\dec{}&10.0~\dec{}  &0.0~\dec{}& 0.0~\dec{}  \\
          & GPT4-0125 &34.8& 29.2  &15.7~\dec{}& 12.4~\dec{}  &43.0&33.3  &20.0~\dec{}& 10.0~\dec{}  &16.7~\dec{}& 6.7~\dec{}  \\
          & DeepSeek &19.1& 14.6  &14.6~\sdec{}& 9.0~\dec{}  &13.3& 16.7  &13.3~\equa{}& 10.0~\dec{}  &10.0~\sdec{}&3.3~\dec{}   \\
          \cmidrule{2-12}
          & \textbf{Average} & 26.6 & 20.8 &  15.2~\dec{42.9\%} &10.1~\dec{51.4\%} &29.9 &26.7  & 17.5~\dec{41.5\%} & 10.0~\dec{62.5\%} &14.2~\dec{52.5\%} & 3.3~\dec{87.6\%}\\
    \midrule
    \multicolumn{2}{c}{\textbf{Average}} & 36.9 & 21.1 & 27.5~\sdec{25.5\%} & 15.8~\sdec{25.1\%} & 36.3 & 18.4 & 29.7~\sdec{18.2\%} & 12.5~\dec{32.1\%} & 20.4~\dec{43.8\%} & 6.9~\dec{62.5\%} \\
    \bottomrule
    \end{tabular}%
    \begin{tablenotes}
    \footnotesize
        \item[1] \sinc{} / \sdec{} means that the pass rate has increased / decreased by less than 30\% (ratio). \inc{} / \dec{} means that the pass rate has increased / decreased by more than 30\% (ratio) .  
        \item[2] The change magnitudes are calculated by comparing to "Original" and the "Original(Structure)". 
    \end{tablenotes}
    \end{threeparttable} 
    }
  \label{tab:Generation}%
   \vspace{-2mm}
\end{table*}%

\begin{table*}[t]
  \centering
  \caption{\textbf{The performance of LLMs in code completion scenarios.} "Original" means the pass rate of all raw code tasks. "Original(Semantic)" means the pass rate of those raw code tasks which later perform semantic obfuscation. All the numbers have omitted the percentage sign.}
  \resizebox{0.999\linewidth}{!}{
    \begin{threeparttable} 
    \begin{tabular}{cccccc|cccccc}
    \toprule
    \multirow{2}[4]{*}{\textbf{Software}} & \multirow{2}[4]{*}{\textbf{Model}} & \multicolumn{2}{c}{\textbf{Original}} & \multicolumn{2}{c|}{\textbf{Symbol}} & \multicolumn{2}{c}{\textbf{Original (Semantic)}} & \multicolumn{2}{c}{\textbf{Semantic}} & \multicolumn{2}{c}{\textbf{Symbol+Semantic}} \\
\cmidrule{3-12}          &       & \textbf{CPR} & \textbf{TPR} &\textbf{CPR} & \textbf{TPR} & \textbf{CPR} & \textbf{TPR} & \textbf{CPR} & \textbf{TPR} & \textbf{CPR} & \textbf{TPR} \\
    \midrule
    \multirow{5}[2]{*}{redis} 
          & GPT3.5 &19.0&9.0 &30.9~\inc{}& 11.5~\sinc{}&17.9& 13.2&13.2~\sdec{}&7.5~\dec{} &36.4~\inc{}& 11.2~\sdec{}  \\
          & GPT4-1106 &41.5&25.4&32.9~\sdec{}&14.2~\dec{}&51.9&36.8 &44.3~\sdec{}&25.5~\dec{} &39.6~\sdec{}&20.7~\dec{}   \\
          & GPT4-0125   &39.3&15.4 &36.1~\sdec{}&15.9~\sinc{}&43.4&17.9 &38.7~\sdec{}& 24.5~\inc{}  &49.1~\sinc{}& 25.5~\inc{}   \\
          & DeepSeek & 40.6 &24.8 & 30.9~\sdec{} &4.0~\dec{} &54.7 &40.6 & 56.6~\sinc{}  &31.1~\sdec{}  & 53.8~\sdec{}&32.1~\sdec{} \\
          \cmidrule{2-12}
          & \textbf{Average} & 35.1 & 18.7 &  32.7~\sdec{6.8\%} &11.4~\dec{39.0\%} &42.0 &27.1  & 38.2~\sdec{9.0\%} & 22.2~\sdec{18.1\%} &44.7~\sinc{6.4\%} & 22.4~\sdec{17.3\%}\\
    \midrule
    \multirow{5}[2]{*}{libvips} 
          & GPT3.5   &19.7&  8.4  &31.0~\inc{}&  13.3~\inc{} &41.2& 11.8   &29.4~\sdec{}& 5.9~\dec{}   &41.2~\equa{}& 10.6~\sdec{}  \\
          & GPT4-1106 & 36.0 & 22.2  & 31.5~\sdec{} & 18.2~\sdec{}  & 35.3 &23.5 & 41.1~\sinc{}&23.5~\equa{} &35.3~\equa{}&17.7~\sdec{} \\
          & GPT4-0125 & 44.3 & 27.1  & 33.5~\sdec{} & 22.2~\sdec{}  & 47.0&29.4 & 47.0~\equa{} &23.5~\sdec{} &29.4~\dec{}&29.4~\equa{} \\
          & DeepSeek & 36.9&  25.1  &37.0~\sinc{}&  21.7~\sdec{} &41.1&  23.5  &41.1~\equa{}& 23.5~\equa{} &35.3~\sdec{}&23.5~\equa{} \\
          \cmidrule{2-12}
          & \textbf{Average} & 34.2&20.7&  33.3~\sdec{2.6\%} &18.9~\sdec{8.7\%}& 41.2 &22.1&39.7~\sdec{3.6\%}  &19.1~\sdec{13.6\%} &35.3~\sdec{14.3\%}&20.3~\sdec{8.1\%}\\
    \midrule
    \multirow{5}[2]{*}{lvgl} 
          & GPT3.5 &  22.4 &18.2 & 27.2~\sinc{}  & 20.5~\sinc{}  &6.7 & 0.0 & 6.7~\equa{}  & 0.0~\equa{} &6.7~\equa{} &0.0~\equa{} \\
          & GPT4-1106 & 39.6  &30.0 & 31.4~\sdec{}  &24.1~\sdec{} & 20.0 & 13.3 & 20.0~\equa{}  &13.3~\equa{} &13.3~\dec{} &13.3~\equa{}  \\
          & GPT4-0125 & 44.2  &33.00  & 36.3~\sdec{} &27.1~\sdec{} &20.0 &13.3 & 20.0~\equa{}  &13.3~\equa{} &13.3~\dec{} &6.7~\dec{} \\
          & DeepSeek & 35.2  & 29.4  &28.2~\sdec{}  &11.9~\dec{}  & 20.0 &13.3  & 20.0~\equa{}  &13.3~\equa{}  &13.3~\dec{} &13.3~\equa{} \\
          \cmidrule{2-12}
          & \textbf{Average} & 35.4  &27.7 & 30.8~\sdec{13.0\%}  &20.9~\sdec{24.5\%} & 16.7  &10.0 & 16.7~\equa{}  &10.0~\equa{}&11.7~\sdec{29.9\%}&8.3~\sdec{17.0\%}\\
    \midrule
    \multirow{5}[2]{*}{libgits} 
          & GPT3.5    &15.4& 12.8  &15.4~\equa{}& 12.8~\equa{}  &10.0& 10.0   &10.0~\equa{}& 10.0~\equa{}   &20.0~\inc{}& 10.0~\equa{}  \\
          & GPT4-1106  &10.3& 10.3  &6.4~\dec{}&6.4~\dec{} &60.0& 30.0    &50.0~\sdec{}& 30.0~\equa{}   &20.0~\dec{}& 20.0~\dec{}  \\
          & GPT4-0125  &12.8& 10.3  &7.7~\dec{}&7.7~\sdec{} &50.0& 40.0  &50.0~\equa{}& 30.0~\sdec{}  &40.0~\sdec{}&30.0~\sdec{}   \\
          & DeepSeek  &25.6&  21.8  &23.0~\sdec{}& 19.2~\sdec{}  &30.0&30.0  &40.0~\inc{}& 20.0~\dec{}   &30.0~\equa{}& 20.0~\dec{}  \\  
          \cmidrule{2-12}
          & \textbf{Average} & 16.0  &13.8 &13.1~\sdec{18.1\%}  & 11.5~\sdec{16.7\%} &37.5 &27.5& 37.5~\equa{}& 22.5~\sdec{18.2\%}& 27.5~\sdec{26.7\%}&20.0~\sdec{27.3}\\
    \midrule
    \multirow{5}[2]{*}{fluent} 
          & GPT3.5 &11.2 &7.87 &16.9~\inc{} &11.2~\inc{} &0.0&0.0  &0.0~\equa{}& 0.0~\equa{}  &18.2~\inc{}&0.0~\equa{}  \\
          & GPT4-1106 &25.8 &23.6 &16.9~\dec{} &11.2~\dec{} &9.1& 9.1  &18.2~\inc{}& 9.1~\equa{}   &9.1~\equa{}& 9.1~\equa{} \\
          & GPT4-0125  &30.3 &27.0 &20.2~\dec{} &19.1~\sdec{} &18.2& 18.2 &18.2~\equa{}& 9.1~\dec{}   &18.2~\equa{}& 9.1~\dec{}  \\
          & DeepSeek &21.4  &18.0 &14.6~\dec{} &11.2~\dec{} &18.2&18.2  &18.2~\equa{}&18.2~\equa{}  &9.1~\dec{}&9.1~\dec{}   \\
          \cmidrule{2-12}
          & \textbf{Average} & 22.2 & 19.1& 17.2~\sdec{22.5\%}  &13.2~\dec{30.9\%} & 11.4  & 11.4 & 13.7~\sinc{20.2\%} &9.1~\sdec{20.2\%} &13.7~\sinc{20.2\%} &6.8~\dec{40.4\%}\\
    \midrule
    \multicolumn{2}{c}{\textbf{Average}} & 28.6 & 20.0 & 25.4~\sdec{11.2\%} & 15.2~\sdec{24.0\%} & 29.7 & 19.6 & 29.1~\sdec{2.0\%} & 16.6~\sdec{15.3\%} & 26.6~\sdec{10.4\%} & 15.6~\sdec{20.4\%} \\
    \bottomrule
    \end{tabular}%
    \begin{tablenotes}
    \footnotesize
    \item[1] \sinc{} / \sdec{} means that the pass rate has increased / decreased by less than 30\% (ratio) . \inc{} / \dec{} means that the pass rate has increased / decreased by more than 30\% (ratio) .  
    \item[2] The change magnitudes are calculated by comparing to "Original" and the "Original(Semantic)". 
    \end{tablenotes}
    \end{threeparttable} 
    }
  \label{tab:Completion}%
   \vspace{-2mm}
\end{table*}%

In this section, we describe our experimental setup and the evaluation results of four LLMs on our dataset. Our evaluation focuses primarily on the performance of LLMs on our dataset, and whether code obfuscation can further reveal the true capabilities of these LLMs.

\subsection{Evaluation Setup}

\subsubsection{\textbf{Model selection}}
We chose a general-purpose large language model, ChatGPT, and a code-focused large language model, DeepSeek. Both of them are mature and widely-used LLMs. For ChatGPT, we use the “gpt-3.5-turbo-1106”~\cite{gpt3.5}, "gpt-4-turbo-1106"~\cite{gpt4.0}, and "gpt-4-turbo-0125"~\cite{gpt4.0} in our experiments.
For DeepSeek, we use DeepSeek-coder-v2 \cite{deepseek-coder} with the default settings. We use default value for LLM's parameters when generating code.

\subsubsection{\textbf{Raw data selection}}
\label{sec:data}
Note that the training datasets for gpt-4-turbo-1106 were finalized as of April 2023 (gpt-3.5-turbo-1106 is also before that time). 
We selected the code starting from May 2023 \checked{to Dec 2023} as the original data. Therefore, for gpt-3.5-turbo-1106 and gpt-4-turbo-1106, these codes were not included in the training set, while for gpt-4-turbo-0125 and DeepSeek-coder-v2, they might have encountered some of these code snippets during training. By doing so, we not only ensure a balanced distribution in our dataset (with both seen and unseen data), but also objectively and authentically demonstrate the generative capabilities of LLMs across different types of code.

\subsubsection{\textbf{Evaluation Metrics}}
To assess the correctness of the generated code snippet, we employed two key performance metrics to measure the code generation capabilities of the LLMs in real-world development scenarios: \textbf{Compile Pass Rate (CPR)} and \textbf{Test Pass Rate (TPR)}. We first replace the original function with the function generated by the LLMs and then compile the software. After that, we perform system tests associated with that function.

CPR and TPR are representative metrics that can illustrate generated code that passes compilation and tests respectively. CPR demonstrates the basic ability of large language models to generate syntactically correct code, while TPR reflects the capability of large language models to understand and correctly generate complex functional code. Specifically, TPR can demonstrate whether the LLM can be used directly in production scenarios. Both CPR and TPR use pass@5 rate to eliminate fluctuations.

\subsection{Results and Analysis}
\label{sec:experiments}

To evaluate the performance of large models on our dataset, we conduct experiments on four large models using \datasetO{}. Specifically, we explore the effectiveness of code obfuscation and the various among different obfuscation strategies. We investigate the following three research
questions:
\begin{itemize}[leftmargin=*, topsep=1pt, itemsep=1pt]
    \item \noindent\textbf{RQ1:} How effective are large language models in generating code on our datasets?
    \item \noindent\textbf{RQ2:} How does code obfuscation further reveal the capabilities of LLMs, and how effective are different obfuscation strategies?
    \item \noindent\textbf{RQ3:} What are the issues hidden in LLM-generated code?
\end{itemize}

\subsubsection{\textbf{RQ1: How effective are large language models in generating code on our datasets}}

The model usually performs better on the code similar to training data. Thus, we collected code data after model training and utilized code obfuscation methods to further eliminate code leakage. In this research question, we conduct experiments with several widely used and proven effective large language models on our collected dataset. The overall results are presented in Table~\ref{tab:Generation} and Table~\ref{tab:Completion}.

The CPRs are among \checked{16.0\% to 49.0\%} on original raw code, with an average of \checked{32.8\%}, which is significantly lower than the LLM's performance on traditional datasets (e.g., humaneval). The average TPRs are \checked{21.1\%} and \checked{20.0\%} on original code generation and completion. After we applied different levels of code obfuscation, this result dropped to \checked{6.9\% and 15.6\%}. This indicates that the code generated by LLMs is difficult to pass the official tests of software and be directly used in production environments. Developers still need to manually fix and adjust the code.

\vspace{-1mm}
\begin{tcolorbox}[boxrule=1pt,boxsep=1pt,left=2pt,right=2pt,top=2pt,bottom=2pt]
\small
\textcolor{red}{\faIcon{user-edit}} \noindent\textbf{Conclusion for RQ1:} Large Language Models still fall significantly short of meeting the requirements for unfamiliar code generation/completion tasks in real-world production environments. The basic pass rate of compilation remains a significant bottleneck. Even if it passes the compilation, it may still not meet the actual functional requirements of the software. 
\end{tcolorbox}

\subsubsection{\textbf{RQ2: How does code obfuscation further reveal the capabilities of LLMs, and how effective are different obfuscation strategies}}

\begin{figure*}[t]
    \centering
    \begin{subfigure}[b]{0.245\textwidth}
        \centering
        \includegraphics[width=\textwidth]{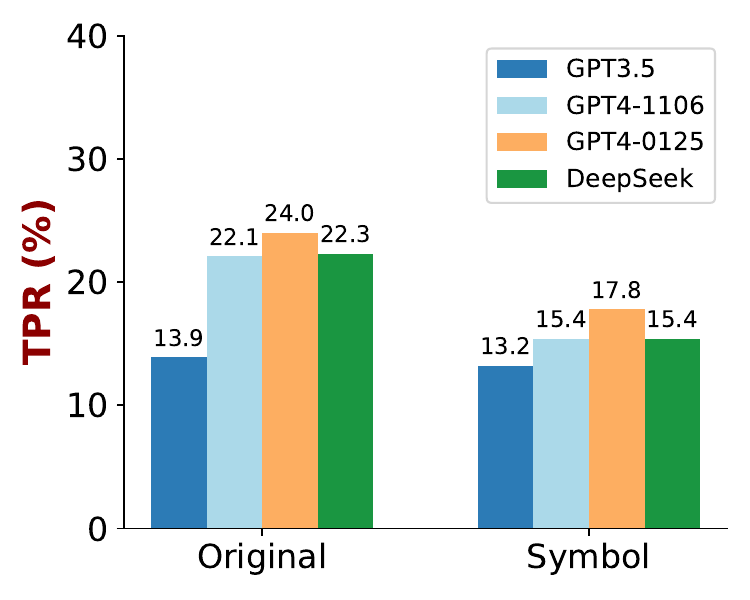}
        \caption{symbol obfuscation}
        \label{fig:sub1}
    \end{subfigure}
    \hfill
    \begin{subfigure}[b]{0.37\textwidth}
        \centering
        \includegraphics[width=\textwidth]{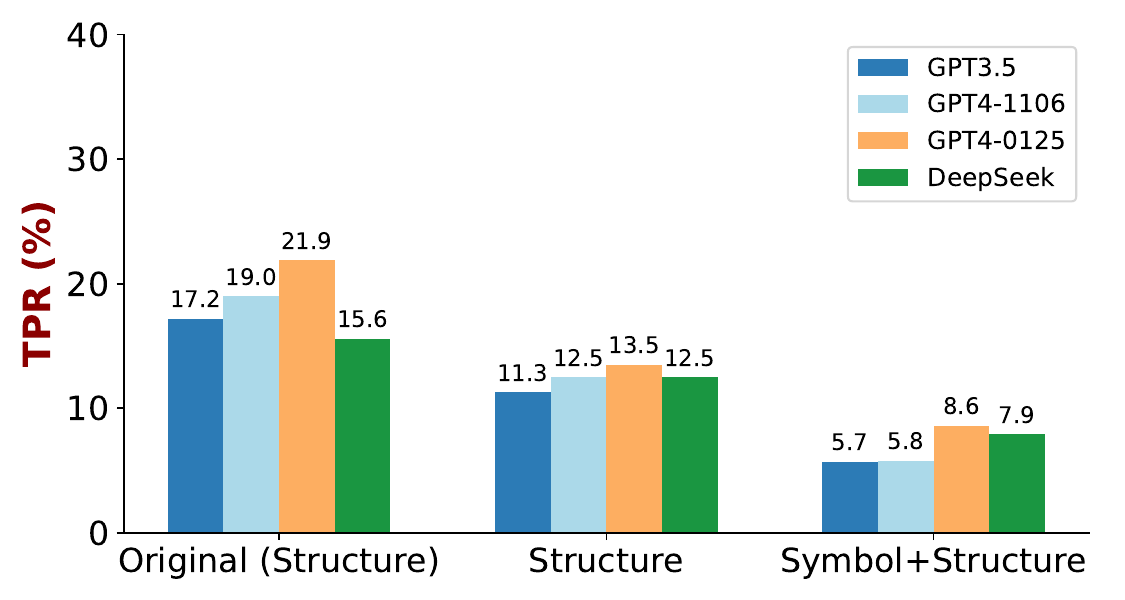}
        \caption{Stucture(\&symbol) obfuscation}
        \label{fig:sub2}
    \end{subfigure}
    \hfill
    \begin{subfigure}[b]{0.37\textwidth}
        \centering
        \includegraphics[width=\textwidth]{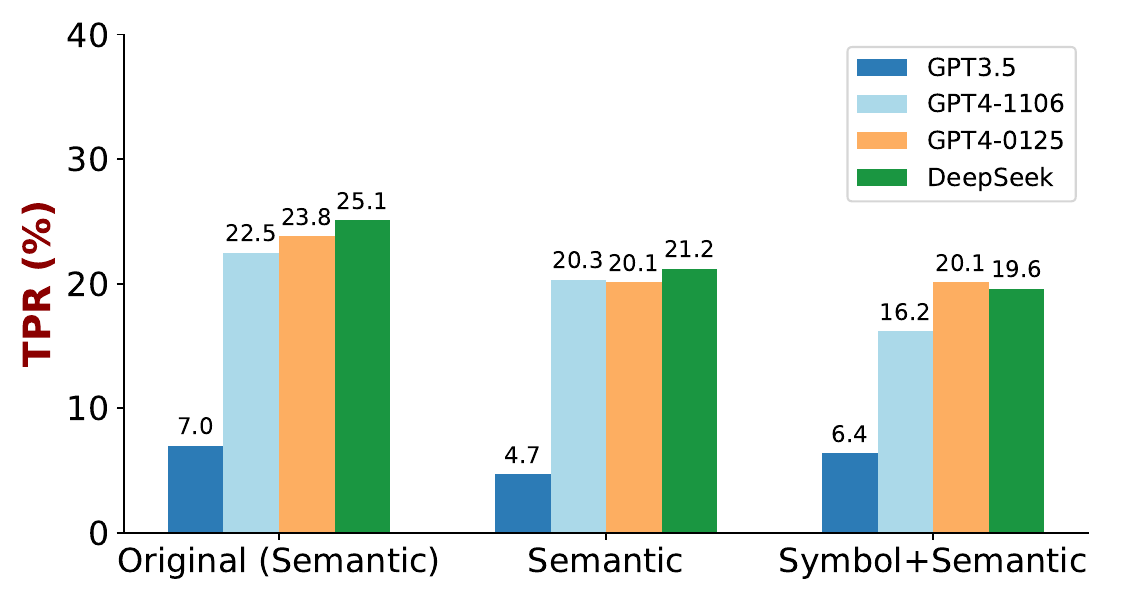}
        \caption{Semantic(\&symbol) obfuscation}
        \label{fig:sub3}
    \end{subfigure}
    \caption{TPR under different code obfuscation strategies. \textbf{After introducing various degrees of obfuscation, the rankings of four LLMS tend to be stabilize and consistent.}}
    \label{fig:TPR}
    \vspace{-2mm}
\end{figure*}

We use three obfuscation strategies (Section~\ref{Sec4.2}) to modify the collected raw code to minimize the possibility that the code might resemble the training data.
We specifically compare the TPR before and after different code obfuscations to evaluate the syntactic and functional correctness of generated code. We try them separately and also their combinations. The results are shown in \figurename~\ref{fig:TPR} (detailed data can be found in Table~\ref{tab:Generation} and Table~\ref{tab:Completion}).

\textbf{Overall effect.}
All obfuscation strategies have reduced the TPR of code generation by LLMs. Since we did not deliberately increase the difficulty and complexity of the code in code obfuscation (discussed in Section~\ref{sec:threat}), code obfuscation can make the tested code less familiar to large models. Even though the decline may not be particularly significant, it represents the trend of the LLMs' capabilities, and if applied in a large-scale production environment, it could still have a considerable impact.

As for tested LLMs, GPT4-1106, GPT4-0125 and DeepSeek Code are obviously better than GPT3.5. However, the ranking of their capability on the Original code is inconsistent. After obfuscation, their ranking becomes more stable (GPT4-0125\textgreater DeepSeek Code\textgreater GPT4-1106\textgreater GPT3.5). This further proves that using code obfuscation can more accurately demonstrate the capabilities of the models.

\textbf{Strategy comparison.}
In this paper, we intuitively determine the effectiveness of code obfuscation strategies by the TPR decrease ratio comparing to the Original code tasks (eliminate the inflated capabilities of the large model). Symbol and structure obfuscation can be very effective with the average TPR decrease ratio of \checked{24.6\% and 32.1\%}. Semantic obfuscation is not that effective, with an average decrease ratio \checked{15.3\%}. This is mainly because our current semantic obfuscation strategies are still relatively simple (using some heuristics to rewrite code manually), without making in-depth modifications to the code. We also find that the use of mixed strategies can lead to a further decrease in TPR (Symbol + Structure can have an average decrease ratio of \checked{62.5\%}), demonstrating the effectiveness of mixed strategies.

Although different obfuscation methods have different effectiveness, we still recommend that future researchers should try various obfuscation strategies and their combinations, which can not only ensure the fairness of code obfuscation but also explore the performance of LLMs on more types of code implementations.

\vspace{-1mm}
\begin{tcolorbox}[boxrule=1pt,boxsep=1pt,left=2pt,right=2pt,top=2pt,bottom=2pt]
\small
\textcolor{red}{\faIcon{user-edit}} \noindent\textbf{Conclusion for RQ2:} All obfuscation strategies can help eliminate the inflated code generation capabilities of LLMs. symbol and structure obfuscation are particularly effective. Future research could further explore more sophisticated semantic obfuscation methods and additional strategies, and automate the entire obfuscation process efficiently.
\end{tcolorbox}

\subsubsection{\textbf{RQ3: What are the issues hidden in LLM-generated code}}
\label{sec:common issues}

To enable developers to better utilize the code generated by large language models for downstream development tasks, we further analyzed common issues hidden in the generated code that may affect development tasks.
Due to the large size of the \datasetO{}, we employed stratified sampling for manual code review. 
The sample size was calculated through the finite population correction, ensuring our sample accurately represents the dataset. 
We set the confidence level to 95\% and the margin of error to 5\%, which are standard statistical thresholds.
Through this method, we extracted 299 pieces of LLM-generated code from five software and identified the following three categories of issues:
\begin{itemize}[leftmargin=*, topsep=1pt, itemsep=1pt]
  \item \textbf{Syntax errors:} Codes lead to compilation errors.
  \item \textbf{Functional errors:} Codes fail to meet the functional requirements of development (fail the official tests).
  \item \textbf{Non-functional code quality issues:} Codes potentially causing performance and reliability issues.
\end{itemize}
Due to the target software being from diverse domains, the functionalities and the code semantics are various, so we mainly focus on \textbf{syntax errors} and \textbf{non-functional code quality issues}.

\begin{table}[t]
  \centering
  \caption{Syntax errors of LLM-generated code}
  \vspace{-1mm}
  \begin{tabularx}{0.49\textwidth}{Xr} 
    \toprule
    \textbf{Category} & \textbf{Proportion} \\
    \midrule
    \textbf{Function and Type Declaration Errors} & \textbf{45.24\%} \\
    \hspace{2em}Implicit declaration of function & 30.56\% \\
    \hspace{2em}Type conflict & 13.35\% \\
    \hspace{2em}API parameter count mismatch & 1.02\% \\
    \hspace{2em}Undeclared type & 0.31\% \\
    \midrule
    \textbf{Data Structure and Member Access Errors} & \textbf{26.70\%}\\
    \hspace{2em}Non-existent structure member & 19.87\% \\
    \hspace{2em}Misuse of structure pointer & 6.60\% \\
    \hspace{2em}Use $\rightarrow$ operator to access an integer member& 0.23\% \\
    \midrule
    \textbf{Type Conversion and Assignment Errors} & \textbf{12.82\%}\\
    \hspace{2em}Making a pointer from an integer without a cast & 8.57\% \\
    \hspace{2em}Incompatible pointer type & 1.25\% \\
    \hspace{2em}Incompatible type assignment & 1.50\% \\
    \hspace{2em}Redefinition & 1.50\% \\
    \midrule
    \textbf{Scope and Definition Errors} &\textbf{1.89\%} \\
    \hspace{2em}Conflict between static and non-static declarations & 1.50\% \\
    \hspace{2em}Incorrect access to structure or union member & 0.39\% \\
    \midrule
    \textbf{Other Syntax Errors} &\textbf{13.35\%} \\
    \hspace{2em}Lvalue required as the left operand of assignment & 6.60\% \\
    \hspace{2em}Incorrect use of array, pointer, or vector & 1.73\% \\
    \hspace{2em}Assignment to expression with array type & 0.63\% \\
    \hspace{2em}Incorrect use of parentheses & 0.63\% \\
    \hspace{2em}Invalid binary operands & 0.55\% \\
    \hspace{2em}Expected expression error & 0.47\% \\
    \hspace{2em}Array subscript is not an integer & 0.23\% \\
    \hspace{2em}Subscripted value is pointer to function & 0.23\% \\
    \hspace{2em}Others & 2.28\% \\
    \bottomrule
  \end{tabularx}
  \label{tab:errorCategories}
   \vspace{-2mm}
\end{table}

\textbf{Syntax Errors.}
We refer to the LLM-generated code that causes compilation errors in the software as syntax error code. 
Such errors significantly impact development efficiency.
We extract error logs and manually analyze the code that failed to compile, ultimately categorizing the syntax errors into 5 major categories and 21 subcategories, as shown in Table~\ref{tab:errorCategories}.

In summary, large language models perform worst in handling function and type declarations, particularly implicit function declarations and type conflicts. And LLMs perform poorly in generating code related to structures, often resulting in issues such as accessing non-existent members.
These errors may be due to the presence of similarly named functions or structures in the training data of the LLMs, leading to the use of functions or structures outside the provided context in code generation tasks.
We recommend that developers focus on checking external dependencies such as called functions and structures when using LLMs to generate code. 
Additionally, LLMs often perform poorly when handling code involving pointers, leading to errors such as converting integers to pointers without casting, and incompatible pointer types.

\begin{figure}[t]
\centering
\begin{subfigure}{0.99\columnwidth}
  \centering
  \includegraphics[width=\linewidth]{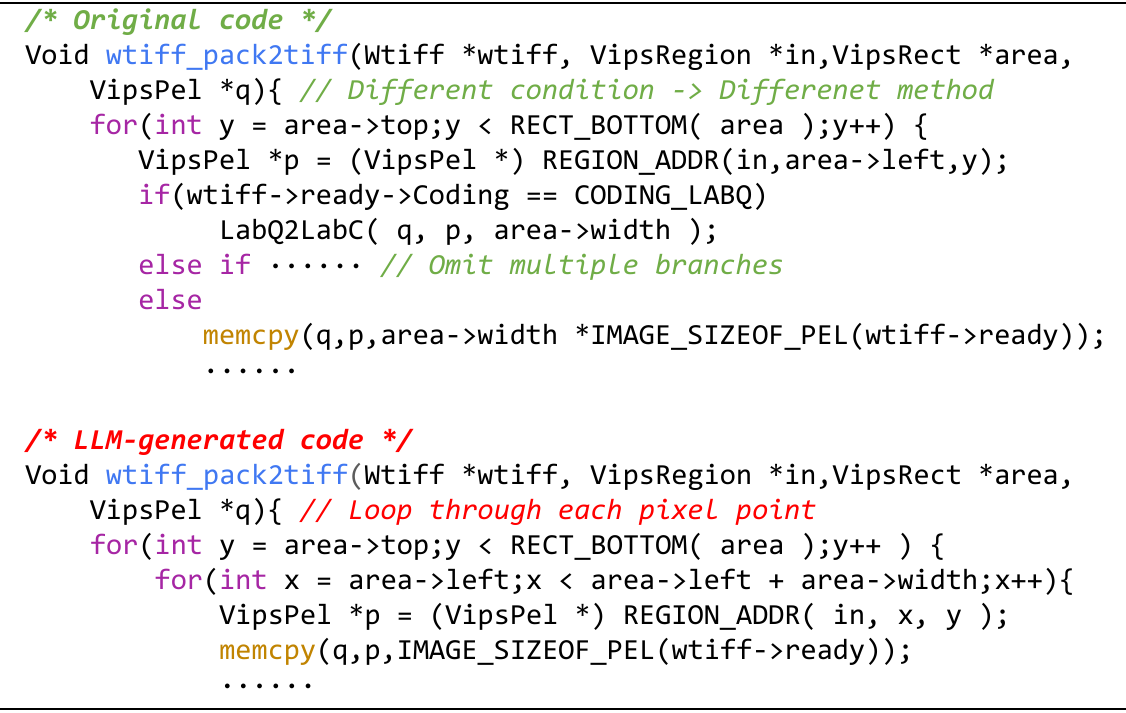}
  \label{fig:RQ3 example1}
  \vspace{-5mm}
  \caption{An example of low efficiency code.}
\end{subfigure}%
\newline 
\begin{subfigure}{0.99\columnwidth}
  \centering
  \includegraphics[width=\linewidth]{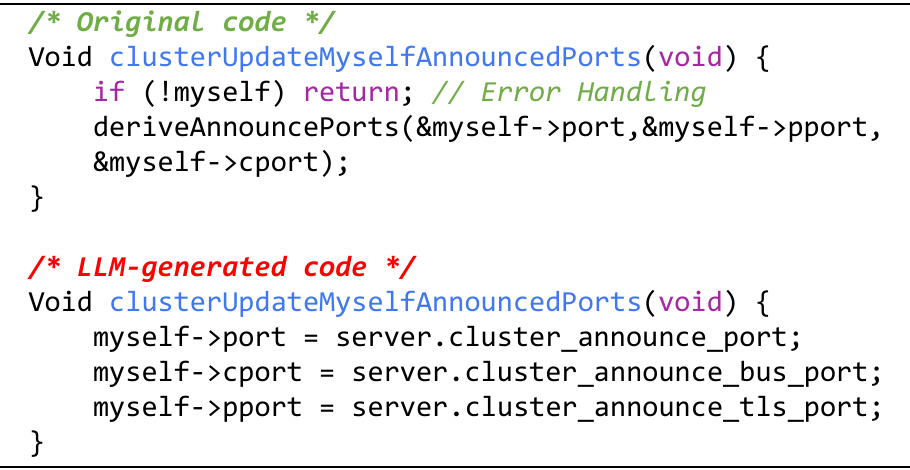}
  \label{fig:RQ3 example2}
   \vspace{-5mm}
  \caption{An example of missing error handling.}
\end{subfigure}
\vspace{-5mm}
\caption{Examples of non-functional code quality issue.}
\label{fig:RQ3}
\vspace{-6mm}
\end{figure}

\textbf{Non-functional code quality issues.}
For LLM-generated codes that have passed systematic testing, we conducted a manual code quality review and found that the code often had non-functional quality issues such as poor performance, and security vulnerabilities.
These issues do not directly affect the normal execution of the code, but they can pose potential threats to the software.
For example, poor performance may cause system response time delays, affecting user experience and overall performance; security vulnerabilities may be exploited by malicious attackers, leading to data breaches or system crashes.
We have summarized common non-functional code quality issues into the following three categories:

\begin{itemize}[leftmargin=*, topsep=1pt, itemsep=1pt]

  \item \textbf{Resource Management:}
  In LLM-generated code, resource management code is often inadequate, especially in terms of memory or file management.
  For example, LLM-generated code often does not free dynamically allocated memory at the end of the program.
  Such code can pass compilation and testing, but it may lead to potential memory leaks, resulting in software performance issues.
  \item \textbf{Code Efficiency:}
  \figurename~\ref{fig:RQ3}(a) shows an example of low code efficiency. 
  \textit{Wtiff\_pack2tiff} is an image data conversion function in the \textit{libvips} software.
  In the original code, the function adopts corresponding processing methods according to different image encoding formats, while the LLM-generated code uses an inefficient loop to process each pixel, modifying and copying pixel by pixel.
  Although the LLM-generated code can meet functionality, its performance is far inferior to the original code, affecting the overall performance of the software.
  \item \textbf{Code Robustness:} The robustness of the generated code is not guaranteed, including missing error checking code, incomplete error handling and inadequate feedback message.

  ~\figurename~\ref{fig:RQ3}(b) shows an example of missing error handling. In the original code, the program first checks if the \textit{myself} pointer is NULL to prevent null pointer reference.
  However, the LLM-generated code does not perform null pointer checks, which could lead to accessing the pointer when it is uninitialized or freed, causing software crashes.
\end{itemize}

\vspace{-1mm}
\begin{tcolorbox}[boxrule=1pt,boxsep=1pt,left=2pt,right=2pt,top=2pt,bottom=2pt]
\small
\textcolor{red}{\faIcon{user-edit}} \noindent\textbf{Conclusion for  RQ3:} Function and Type Declaration Errors are the most common syntax error of LLM generated code.
While, even if the code passes compilation and system testing, it may still have non-functional issues which will negatively affect software reliability and performance. Future research should establish more comprehensive evaluation metrics to evaluate the quality of code generated by LLMs.
\end{tcolorbox} 

\section{Threats to validity}
\label{sec:threat}

\noindent\textbf{Raw data collection.} To simulate real-world development scenarios, we select mature system software from the real world as the target software for our research. We chose these software systems because they are widely used, have a rich development history, and contain mature test suites. We believe our study is representative, although some results may not apply to all kinds of software and all kinds of code language. There are other human efforts evolved in the data collection process (e.g., description rewriting). To minimize the impact of human error, we organize a team of seven senior software engineers, each with at least five years of experience in C programming. Additionally, we conduct a double-check progress in each step. 

\Space{
\noindent\textbf{Description rewriting.} We write functional descriptions for the sampled functions based on understanding the code semantics. For code that already had descriptions, we manually rewrite the descriptions to achieve obfuscation. To minimize the impact of human error on model generation, we organize a team of seven senior software engineers, each with at least 3 years of experience in C programming. Additionally, we conduct a double-check progress on each description.}

\noindent\textbf{Code complexity.} Obfuscation strategies can introduce changes in code complexity, primarily in structural and semantic obfuscation. For example, function inlining in structural obfuscation may lead to an increase in the number of lines of code, potentially affecting the generation capabilities of large models. In practice, we did not intentionally increase the complexity of the code in any obfuscation method; our guiding principle was to ensure that the semantics of the code remained the same before and after obfuscation.

\noindent\textbf{Testing process.} Due to the presence of flaky tests, the testing environment of real software projects can be unstable. Even correct code may fail tests due to contextual or environmental issues. Therefore, for each case, we run tests at least five times to mitigate intermediate test results. Official test suites may not comprehensively test the correctness of functionality, but this represents the best efforts, allowing for a better assessment of whether the code meets development requirements.

\section{Conclusion}
\label{sec:conclusion}

Accurately assessing the code generation capabilities of LLMs is crucial for their evaluation and improvement. While existing works have constructed datasets to gauge these capabilities, three main gaps persist in objectively evaluating LLMs' real potential: the exposure of target code, case timeliness, and dependency availability. These gaps arise because the code in current datasets may have been exposed during the training phase of LLMs, and the continuous training and development of LLMs severely compromise their timeliness. To address the problem, this paper adopts the concept of code obfuscation, altering code at various levels while preserving its functionality and output. We developed a code-obfuscation-based benchmark, \datasetO{}, by collecting \checked{1,354} raw cases from \checked{five} real-world projects, which include function descriptions and code. We then obfuscated descriptions, code, and context dependencies using a three-level strategy (symbol, structure, and semantic). Evaluating \checked{four} LLMs on \datasetO{} and comparing the effectiveness of different obfuscation strategies, we found that after obfuscation, the average test pass rate can decreased by \checked{15.3\%-62.5\%}.

\section*{Acknowledgments}
We thank the anonymous reviewers for their insightful comments. This research was funded by NSFC No.62272473, the Science and Technology Innovation Program of Hunan Province No.2023RC1001, NSFC (No.U2441238, No.62202474) and National University of Defense Technology Research Project No.ZK24-01.

\bibliographystyle{IEEEtran}
\bibliography{mylib}
\end{document}